\documentclass[5p]{elsarticle}

\usepackage{float}
\usepackage{amsmath}
\usepackage{graphicx}
\usepackage{siunitx}
\usepackage{xr-hyper}
\usepackage[colorlinks=true,citecolor=blue,urlcolor=red,linkcolor=red,filecolor=cyan]{hyperref}
\usepackage[capitalise]{cleveref}
\graphicspath{{figs/}} %Setting the graphicspath

\title{Growth of the Solid-Electrolyte Interphase: Electron Diffusion versus Solvent Diffusion}

\author[1,2]{Lukas Köbbing}

\author[1,2,3]{Arnulf Latz}

\author[1,2,3]{Birger Horstmann\corref{cor1}}

\cortext[cor1]{birger.horstmann@dlr.de}

\affiliation[1]{organization={Institute of Engineering Thermodynamics, German Aerospace Center (DLR)},
	addressline={Wilhelm-Runge-Straße 10},
	city={89081 Ulm},
	country={Germany}}
\affiliation[2]{organization={Helmholtz Institute Ulm (HIU)},
	addressline={Helmholtzstraße 11},
	city={89081 Ulm},
	country={Germany}}
\affiliation[3]{organization={Institute of Electrochemistry, Ulm University},
	addressline={Albert-Einstein-Allee 47},
	city={89081 Ulm},
	country={Germany}}

\begin{document}
%\addbibresource{refs.bib}

\begin{abstract}
    The solid-electrolyte interphase (SEI) substantially influences the lifetime of lithium-ion batteries. Nevertheless, the transport mechanism responsible for the long-term growth of the SEI remains controversial. This study aims at discussing the characteristic time and state-of-charge dependence of SEI growth mediated by electron diffusion versus solvent diffusion. We describe both transport mechanisms with continuum models and compare them to experimental results. We show that electron diffusion can explain both the observed state-of-charge dependence and the time dependence. In contrast, we demonstrate that solvent diffusion can reproduce either the state-of-charge dependence or the time dependence of capacity fade. There is no intermediate regime where solvent diffusion can explain both dependencies simultaneously. Furthermore, we emphasize the crucial role of anode voltage and state-of-charge on SEI growth in general. Due to self-discharge, this dependence can explain deviations from the typical square-root behavior in the time domain. We conclude that electron diffusion is the relevant process leading to the state-of-charge dependent SEI growth. Further experiments are needed to investigate the reason for contributions to the capacity fade that are independent of the state-of-charge.
\end{abstract}

\maketitle

%\tableofcontents

%\section*{Keywords}
%\noindent
%- Lithium-ion batteries\\
%- Solid-electrolyte interphase (SEI)\\
%- Battery aging and degradation\\
%- SEI growth\\
%- Diffusion mechanisms\\
%- SOC and time dependence
%
%\section*{Highlights}
%\noindent
%- Discussion of state-of-charge (SOC) dependence and time dependence of SEI growth.\\
%- Electron diffusion describes time and SOC dependence of SEI growth simultaneously.\\
%- Interplay between reaction and diffusion is considered for solvent diffusion.\\
%- Solvent diffusion describes either time or SOC dependence of SEI growth.\\
%- Deviation from square-root behavior in time can be explained with self-discharge.

\section{Introduction}

The solid-electrolyte interphase (SEI) crucially determines the performance and degradation of lithium-ion batteries. Since its first observation, Peled contributed significantly to the theoretical understanding of the SEI: Peled provided the first theoretical description of the SEI in 1979 \cite{Peled1979}, considered its complex structure and composition \cite{Peled1995}, and constructed an equivalent circuit model for a mosaic SEI \cite{Peled1997}. His recent review from 2017 highlights the great importance of ongoing SEI research for the improvement of future batteries \cite{Peled2017}.

Despite a lot of effort that has been going into the theoretical and experimental investigation of the SEI for many years, the SEI is still not fully understood \cite{Winter2009,Wang2018,Horstmann2019}. This results from the complexity of the SEI due to the variety of organic and inorganic SEI components \cite{Aurbach1994,Zhang2005,An2016}, mixed layered and mosaic SEI structures \cite{Aurbach1999,Aurbach2000,Single2016,Li2017,Kolzenberg-Werres2022}, a heterogeneous SEI morphology \cite{Cresce2014,Zheng2014,Bolay2022}, and the contribution of different SEI formation reactions to the evolution of the SEI \cite{Single2017,Single2018,Heiskanen2019,Alzate-Vargas2021,Spotte-Smith2022}.

\begin{figure}[htb]
    % Figure 1
    % width = 1 column
	\centering
	\includegraphics[width=0.45\textwidth]{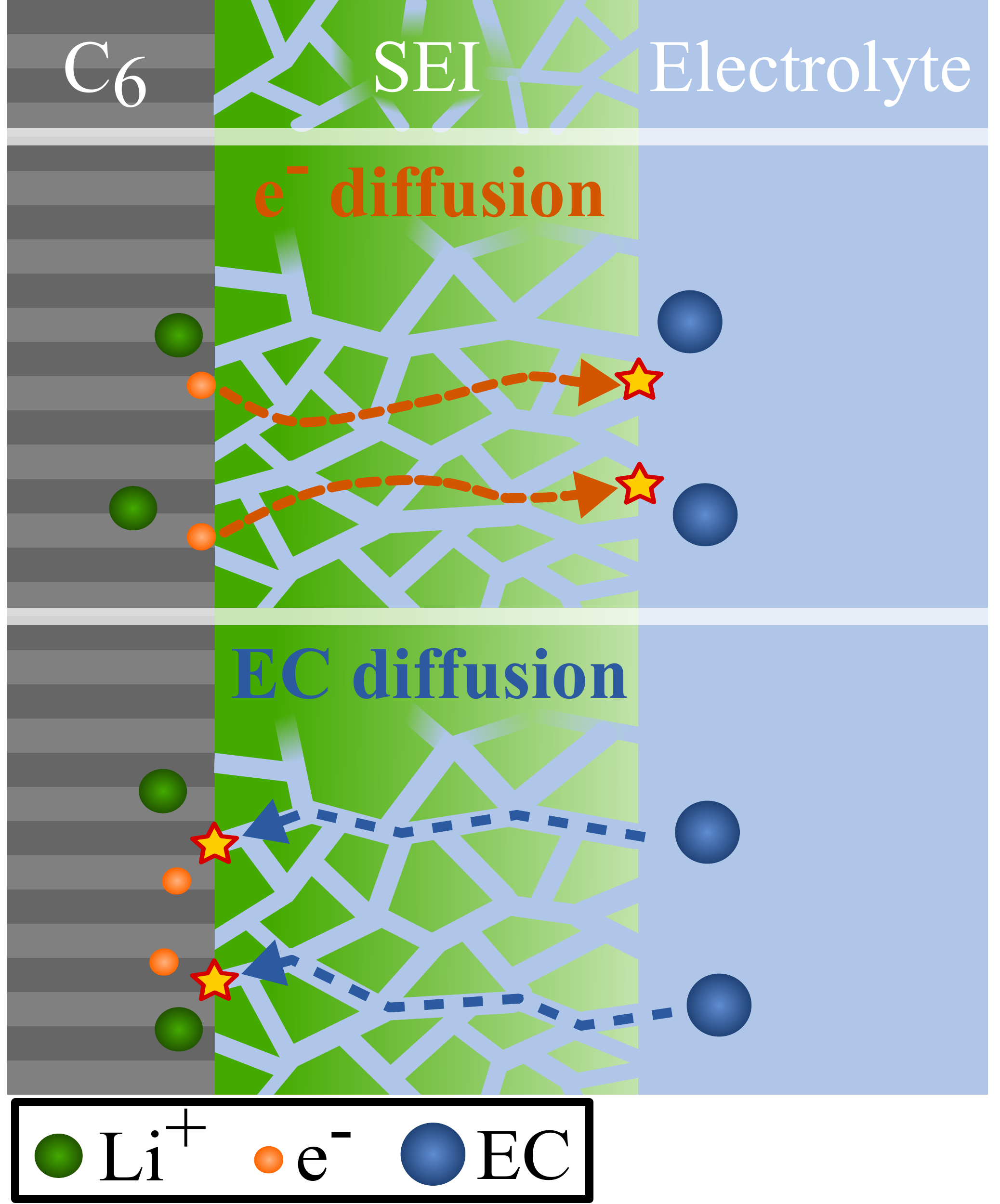}
	\caption{Schematic of the two transport mechanisms investigated in this study. The upper part depicts the diffusion of electrons from the electrode through the SEI towards the electrolyte. The lower part illustrates the diffusion of solvent molecules from the electrolyte through the SEI towards the electrode.}
	\label{fig:scheme}
\end{figure}

One major open question is the exact growth mechanism responsible for the long-term SEI growth. In this paper, we focus on the two most widely used mechanisms: electron diffusion through localized states \cite{Single2017,Single2018,Shi2012,Soto2015,Kolzenberg2020} and solvent diffusion \cite{Single2018,Ploehn2004,Ramasamy2007,Sankarasubramanian2012,Pinson2013,Hao2017}. We illustrate both mechanisms in \cref{fig:scheme}. The most promising mechanism for describing experimental data for SEI growth is electron diffusion \cite{Horstmann2019}, e.g. in form of neutral lithium interstitial diffusion \cite{Single2018,Kolzenberg2020}. However, the literature also provides arguments supporting a porous SEI allowing for solvent diffusion \cite{Krauss2022}. Arguments for solvent diffusion reported in the literature are observed currents from redox shuttles through the SEI \cite{Tang2012,Kranz2019} as well as swelling of the SEI inside the electrolyte compared to a dried SEI \cite{Zhang2022}.

%%%%%%%%%%%%%%%%%%%%%%%%%%%%%%%%%%%%%%%%%%%%%%%%%%%%%%%%%%%
% Figure with label fig:scheme here !!!
%%%%%%%%%%%%%%%%%%%%%%%%%%%%%%%%%%%%%%%%%%%%%%%%%%%%%%%%%%%

To compare these two mechanisms, we investigate the SEI growth during open-circuit battery storage, usually called calendar aging. During calendar aging, the growth of the SEI is the main issue leading to capacity loss of the battery \cite{Keil2016,Keil2017}. Under storage conditions, the literature reports two important observations. The first one is the dependence of capacity loss on the state-of-charge (SOC) \cite{Keil2016,Keil2017,Naumann2018,Werner2021}, and the second one is the square-root behavior in time \cite{Broussely2001,Smith2011,Wang2014,Kolzenberg2022}.

Although SEI growth under storage conditions still lacks a complete understanding, some efforts were made to evaluate the SEI growth during cycling of batteries \cite{Kolzenberg2020,Reniers2019,Das2019}. The experimentally observed heterogeneity of the SEI during cycling was rationalized by electron diffusion \cite{Bolay2022} in contrast to solvent diffusion \cite{Tahmasbi2017}. In fact, even self-discharge during battery storage is a sort of battery operation and influences the growth of the SEI \cite{Single2018}.

To reexamine the usual square-root behavior, Attia et al. investigated the exact behavior of SEI growth in time during storage and cycling \cite{Attia2020}. They provide measurements of capacity loss on carbon black over time during open-circuit storage. The experimental data show a particularly large capacity loss and a clear deviation from the usual square-root behavior in the time domain. This evidences that it is crucial to consider additional effects like self-discharge, at least for large capacity losses.

In this paper, we discuss the theory of SEI growth mediated by electron diffusion and solvent diffusion. We compare the observed dependence of capacity loss on SOC and time to the simulation results. We check first for electron diffusion and second for solvent diffusion whether the two mechanisms can describe both dependencies with the same parameters. For solvent diffusion, we look additionally for an intermediate regime revealing a trade-off between SOC and time dependence. In the last section, we highlight the important role of self-discharge on SEI growth for large capacity losses.

\section{Theory}

We use continuum modeling to describe homogeneous long-term growth of the SEI. From this perspective, the SEI formation consists of a diffusion and a reaction step.
There is a variety of possible SEI formation reactions, but we restrict to a single reaction for simplicity,
\begin{equation}
	2 \mathrm{Li}^{+}+2 \mathrm{e}^{-}+2 \mathrm{EC} \rightarrow \mathrm{Li}_{2} \mathrm{EDC}+\mathrm{R}_\mathrm{g},
	\label{eq:SEI-formation-reaction}
\end{equation}
where lithium ions $\mathrm{Li}^{+}$, electrons $\mathrm{e}^{-}$, and solvent molecules ethylene carbonate ($\mathrm{EC}$) react to lithium ethylene dicarbonate ($\mathrm{Li}_{2} \mathrm{EDC}$) which is often considered as the main SEI component. Here, $\mathrm{R_g}$ is a gaseous byproduct.

To enable the reaction (\ref{eq:SEI-formation-reaction}), all the reagents need to be close together. We consider two possible transport mechanisms to achieve this condition. The first one is the diffusion of electrons through the SEI towards the SEI-electrolyte interface, where the reaction takes place in this case. The second one is the diffusion of solvent molecules through the SEI towards the electrode-SEI interface, where the reaction takes place then. In the following, we derive the SEI growth rate for both diffusion mechanisms.

\subsection{Electron diffusion through localized states}
For the derivation of the theory of electron diffusion in the transport limited regime, we refer to Ref. \cite{Single2018}. This article illustrates the electron diffusion mechanism with the diffusion of neutral lithium interstitials supported by Ref. \cite{Shi2012}. However, as the SEI is not a perfect layer, electrons can diffuse as well via different localized states, for example created by inhomogeneities \cite{Smeu2021}, grain boundaries between different SEI components \cite{Pan2016}, or radicals inside the SEI \cite{Soto2015}.

The capacity loss $Q_{\mathrm{SEI}}$ observed in experiments is related to the thickness of the SEI $L_\mathrm{SEI}$ via
\begin{equation}
	L_\mathrm{SEI}=\frac{v}{s} \frac{Q_{\mathrm{SEI}}}{A F}+L_{\mathrm{SEI},0},
\end{equation}
where $v$ is the mean molar volume of the SEI, $s$ is the mean stoichiometric coefficient of lithium in the SEI formation reaction, $A$ is the surface of the anode, and $F$ is the Faraday constant. The initial capacity loss $Q_{\mathrm{SEI,0}}$ is related to the initial thickness of the SEI by $L_{\mathrm{SEI},0} = vQ_{\mathrm{SEI,0}} / (s A F)$.

According to Ref. \cite{Single2018}, the capacity loss over time due to SEI growth mediated by electron diffusion is given by
\begin{equation}
	\partial_t Q_\mathrm{SEI} = \frac{A^2sF^2D_\mathrm{e^-}}{v} \cdot \frac{c_\mathrm{e^-,0} \cdot e^{-\tilde{U}}}{Q_\mathrm{SEI} + Q_\mathrm{SEI,0}}
	\label{eq:CL-electron-diffusion}
\end{equation}
with $\tilde{U} = FU/R_\mathrm{gas}T$, where $U$ is the anode open circuit potential, $R_\mathrm{gas}$ is the universal gas constant, and $T$ is the temperature. $D_\mathrm{e^-}$ is the diffusion coefficient of electrons through the SEI and $c_\mathrm{e^-,0}$ is the electron concentration at $U=0\mathrm{V}$.

For constant voltage, \cref{eq:CL-electron-diffusion} leads to the well-known square-root behavior in time,
\begin{equation}
	Q_{\mathrm{SEI}}=AF \sqrt{\frac{2s}{v} D_\mathrm{e^-} c_\mathrm{e^-,0}} e^{-\tilde{U}/2} \sqrt{t+t_0}-Q_{\mathrm{SEI,0}},
\end{equation}
where $t_0$ depends on the initial capacity loss such that $Q_{\mathrm{SEI}}$ vanishes at $t=0$.

\subsection{Solvent diffusion}
Next, we need to derive the theory of SEI growth generated by solvent diffusion in detail. In the transport limited regime, solvent diffusion shows no SOC dependence contradicting experimental observations \cite{Single2018}. In contrast, in the reaction limited regime, the time behavior is linear, which does not match experiments as well \cite{Kolzenberg2020}. Therefore, we consider an interplay between the diffusion of solvent molecules and the SEI formation reaction.

The rate of charge consumption due to the SEI formation reaction is described by a Butler-Volmer equation as
\begin{equation}
	R = \frac{j_0}{F} \left[\tilde{c}_\mathrm{EC}(0) e^{-(1-\alpha)\tilde{U}} - e^{\alpha \tilde{U} - \tilde{U}_\mathrm{SEI}}\right].
	\label{eq:reaction-rate}
\end{equation}
Here, $j_0$ is the maximum reaction current, $\tilde{c}_\mathrm{EC}(0)$ is the normalized solvent concentration at the electrode-SEI interface, $\alpha$ is the symmetry factor and $\tilde{U}_\mathrm{SEI} = FU_\mathrm{SEI}/R_\mathrm{gas}T$ is the SEI formation potential with $U_\mathrm{SEI} = 0.8\mathrm{V}$. Typically, the first exponential term in \cref{eq:reaction-rate} describing SEI formation dominates as the SEI product is assumed to be insoluble.

The transport of solvent molecules through the SEI is described by Fickian diffusion,
\begin{equation}
	\frac{\partial \tilde{c}_\mathrm{EC}}{\partial t} = \frac{D_\mathrm{EC}}{L_\mathrm{SEI}^2} \cdot  \frac{\partial ^2 \tilde{c}_\mathrm{EC}}{\partial \tilde{x}^2},
	\label{eq:Fickian-diffusion}
\end{equation}
where $\tilde{c}_\mathrm{EC} = c_\mathrm{EC}/c_\mathrm{EC,bulk}$ is the normalized concentration, $\tilde{x} = x/L_\mathrm{SEI}$ is the distance from the electrode, and $D_\mathrm{EC}$ is the diffusion coefficient of solvent through the SEI.

At the boundaries, the SEI reaction rate $R$ and the bulk concentration, respectively, determine the concentration,
\begin{equation}
    \begin{aligned}
        \quad \frac{D_\mathrm{EC} c_\mathrm{EC,bulk}}{L_\mathrm{SEI}} \cdot \left.\frac{\partial \tilde{c}}{\partial \tilde{x}}\right|_{x=0} = R,\\
	\quad \tilde{c}_\mathrm{EC}(\tilde{x} \rightarrow 1) = 1.\\
    \end{aligned}
    \label{eq:boundary-conditions}
\end{equation} 
From the equations (\ref{eq:reaction-rate}), (\ref{eq:Fickian-diffusion}) and (\ref{eq:boundary-conditions}) we derive the concentration at the electrode-SEI interface as
\begin{equation}
    \tilde{c}_\mathrm{EC}(0) = \frac{1 + \frac{j_0L_\mathrm{SEI}}{FD_\mathrm{EC}c_\mathrm{EC,bulk}}e^{\alpha\tilde{U}-\tilde{U}_\mathrm{SEI}}}{1 + \frac{j_0L_\mathrm{SEI}}{FD_\mathrm{EC}c_\mathrm{EC,bulk}}e^{-(1-\alpha)\tilde{U}}}.
    \label{eq:concentration0}
\end{equation}
The resulting equation describing capacity loss due to SEI growth mediated by solvent diffusion derived from equations (\ref{eq:reaction-rate}) and (\ref{eq:concentration0}) then reads
\begin{equation}
    \partial_t Q_\mathrm{SEI} = Aj_0 \frac{e^{-(1-\alpha)\tilde{U}} - e^{\alpha\tilde{U}-\tilde{U}_\mathrm{SEI}}}{1 + \frac{v j_0}{sAF^2D_\mathrm{EC}c_\mathrm{EC,bulk}}e^{-(1-\alpha)\tilde{U}} \left(Q_\mathrm{SEI} + Q_\mathrm{SEI,0}\right)}.
    \label{eq:CL-solvent-diffusion}
\end{equation}
In the transport limited regime, i.e. $FD_\mathrm{EC} c_\mathrm{EC,bulk}/L_\mathrm{SEI} \ll j_0$, \cref{eq:CL-solvent-diffusion} reveals the square-root behavior in time for constant voltage neglecting the term of the backward reaction,
\begin{equation}
    Q_{\mathrm{SEI}}= AF \sqrt{\frac{2s}{v}D_\mathrm{EC}c_\mathrm{EC,bulk}\left(t+t_0\right)} - Q_\mathrm{SEI,0}.
\end{equation}
In the reaction limited regime, i.e. $j_0 \ll FD_\mathrm{EC} c_\mathrm{EC,bulk}/L_\mathrm{SEI}$, \cref{eq:CL-solvent-diffusion} shows a linear time dependence, but non-trivial SOC dependence,
\begin{equation}
    Q_{\mathrm{SEI}} = Aj_0 \left(e^{-(1-\alpha)\tilde{U}} - e^{\alpha\tilde{U}-\tilde{U}_\mathrm{SEI}}\right) t.
\end{equation}

\section{Results and Discussion}

Now, we evaluate the dependence of capacity fade on SOC and time for open-circuit storage. We analyze the data by Keil et al. \cite{Keil2016}, who measured the capacity loss depending on the SOC for lithium-ion batteries with a nickel cobalt aluminum oxide (NCA) cathode. By measuring the SOC dependence at different times up to 9.5 months, they also provide the time dependence of capacity fade at different states-of-charge. Importantly, they also measured the anode open circuit voltage $U(\mathrm{SOC})$ for the same cells.

We support our analysis with new data for lithium iron phosphate (LFP) cells provided by Naumann et al. \cite{Naumann2018} in the supplementary information. They focus on measuring the time dependence at different states-of-charge but do not provide the anode open circuit voltage curve for the investigated cells. Thus, for this data set, we use the anode open circuit voltage measured for LFP cells in Ref. \cite{Keil2016}.

The authors have already analyzed the SOC dependence in Ref. \cite{Single2018}. Here, we evaluate not only the SOC dependence but also the time dependence. A theory that aims at explaining the observed capacity fade has to describe both dependencies simultaneously.

\subsection{Electron diffusion: SOC and time dependence}

\begin{figure*}[htp]
    % Figure 2
    % width = 2 columns
	\begin{minipage}{0.5\textwidth}
		%\centering
		\includegraphics[width=0.95\textwidth]{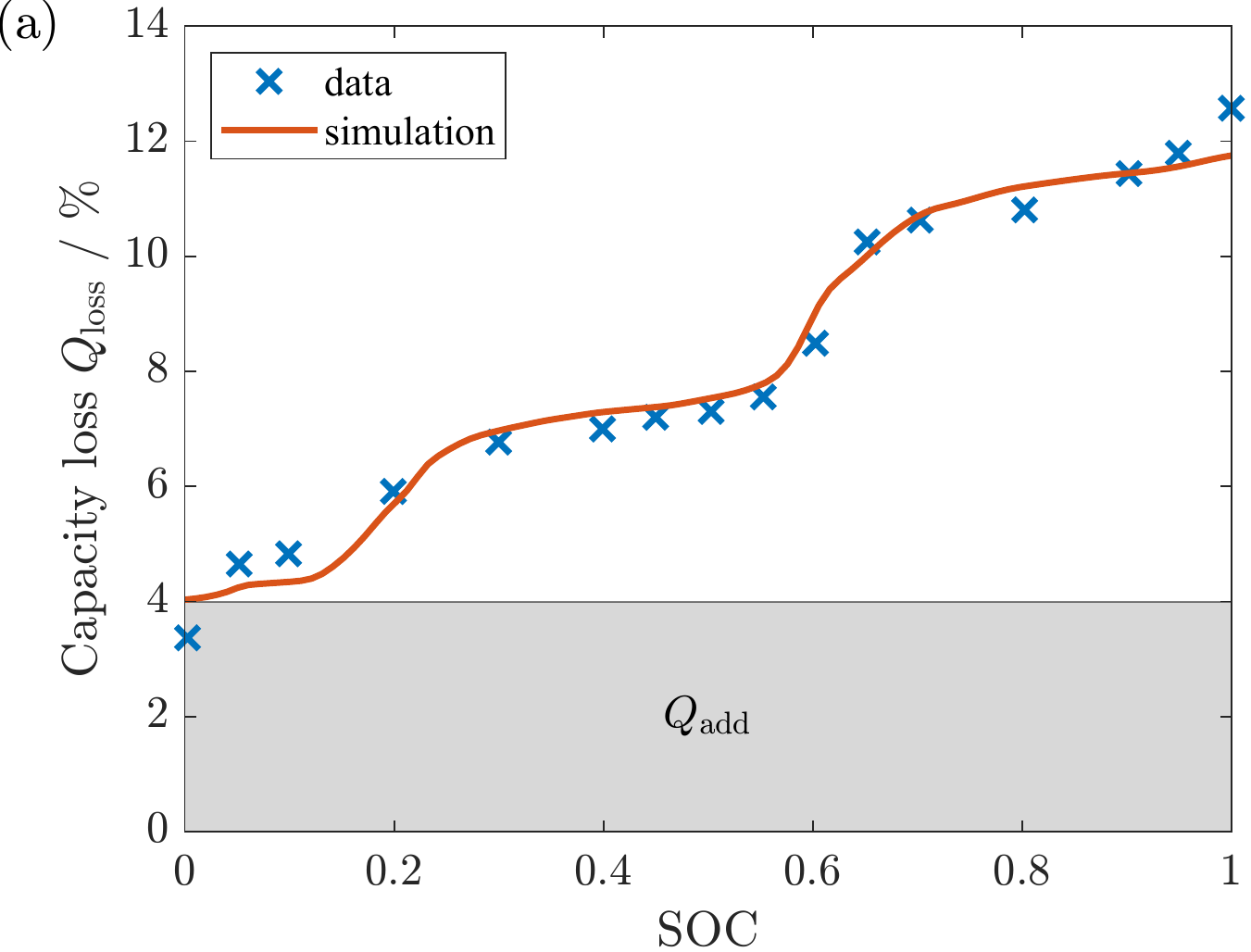}
	\end{minipage}
	\begin{minipage}{0.5\textwidth}
		%\centering
		\includegraphics[width=0.95\textwidth]{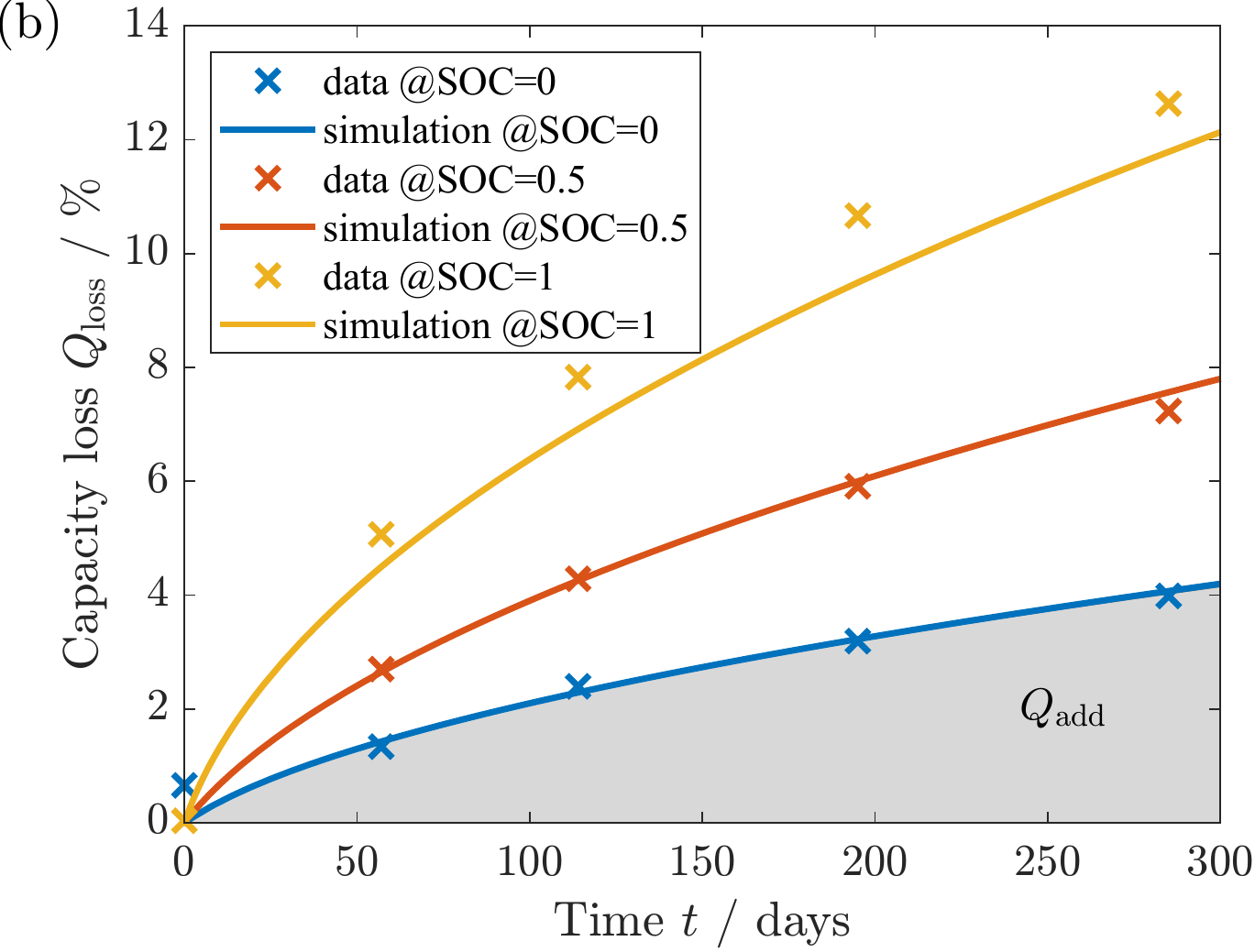}
	\end{minipage}
	\caption{Capacity loss simulated with electron diffusion fitted to the SOC depending data measured in Ref. \cite{Keil2016}. (a) Capacity loss depending on the SOC. The plot shows a good agreement between simulation and data points. (b) Capacity loss depending on time. The agreement is excellent at low and medium SOC, whereas a minor deviation is visible at high SOC.}
	\label{fig:ED-SOC-time}
\end{figure*}

First of all, we want to review the electron diffusion mechanism. The experimentally observed capacity loss $Q_\mathrm{loss}$ consists of a contribution due to SEI growth depending on the SOC and an additional contribution that does not depend on the SOC, i.e. $Q_\mathrm{loss}=Q_\mathrm{SEI}+Q_\mathrm{add}$ \cite{Single2018}. At high SOC, both terms contribute significantly to the total capacity loss. At low SOC, the constant contribution, which serves as an offset for the capacity loss, is the dominant part. We can investigate the time dependence of the SOC depending capacity fade by subtracting the capacity fade observed at $\mathrm{SOC}=0$.

To determine the behavior of capacity loss that is independent of the SOC, we analyze the time dependence of the capacity fade measured at $\mathrm{SOC}=0$. As only a few data points exist for the time dependence in Ref. \cite{Keil2016}, we support our findings with additional data from Ref. \cite{Naumann2018} in the supplementary information. For both data sets, we find that a square-root behavior in time agrees well with the data points (see \ref{fig:CL-time-SOC0} and \ref{fig:CL-time-SOC0-naumann}). Due to the initial SEI thickness, we observe a time offset in the square root that can be related to the initial capacity loss $Q_\mathrm{SEI,0}$. Furthermore, the anode overhang area leads to the observation of an increase in the capacity at the beginning of the experiment at low SOC \cite{Naumann2018}. We account for this with a negative capacity loss at $t = 0$. For further use, we shift the data such that the curve describing the capacity loss vanishes initially, i.e. it starts at the origin.

The cause of the capacity loss at $\mathrm{SOC} = 0$ is not known. Possible reasons are SEI growth during the check-up cycles or mediated by a different transport mechanism independent of the SOC. In addition, other degradation than SEI growth, like particle fracture and contact loss, can cause the observed capacity fade.

After discussing the SOC independent capacity loss, we want to analyze the SOC dependence and the time dependence of the capacity fade now. For this, we subtract the curve obtained at $\mathrm{SOC}=0$ from the data points to investigate the time dependence of the SOC depending contribution to the capacity fade. We use our model in \cref{eq:CL-electron-diffusion} with the determined $Q_\mathrm{SEI,0}$. The anode voltage $U$ depends on the SOC as shown in Ref. \cite{Keil2016}, where the SOC changes during storage due to capacity loss and capacity retention after check-ups. The constant factor in \cref{eq:CL-electron-diffusion}, namely $A^2sF^2D_\mathrm{e^-}c_\mathrm{e^-,0}/v$, serves as a single fitting variable. In the following, we examine the data provided by Keil et al. \cite{Keil2016} and analyze the data by Naumann et al. \cite{Naumann2018} in the supplementary information (see \ref{fig:ED-SOC-time-naumann}).

The fit of the electron diffusion model equation (\ref{eq:CL-electron-diffusion}) to the capacity loss depending on the SOC is shown in \cref{fig:ED-SOC-time}(a). Overall, the simulation agrees well with the experimental data, i.e. electron diffusion can explain the SOC dependence of capacity fade. Only at very low and very high SOC, we observe major deviations that may account for additional effects.

%%%%%%%%%%%%%%%%%%%%%%%%%%%%%%%%%%%%%%%%%%%%%%%%%%%%%%%%%%%
% Figure with label fig:ED-SOC-time here !!!
%%%%%%%%%%%%%%%%%%%%%%%%%%%%%%%%%%%%%%%%%%%%%%%%%%%%%%%%%%%

With the same parameters, in \cref{fig:ED-SOC-time}(b) we compare the time dependence to the experiment. We observe an excellent agreement between the simulated time dependence and the data at low and medium SOC. The data as well as the simulation approximately show a square-root behavior in time. At high SOC, the deviation between simulation and data is more pronounced, indicating an additional contribution to capacity loss at very high SOC. This is in accordance with the deviation at high SOC observed in the SOC dependence.

Altogether, we have shown that electron diffusion is able to describe the SOC dependence and the time dependence with the same parameters. Thus, electron diffusion can explain the observed behaviors of SEI growth quite accurately.

\subsection{Solvent diffusion: SOC and time dependence}

\begin{figure*}[tb]
    % Figure 3
    % width = 2 columns
	\begin{minipage}{0.5\textwidth}
		\includegraphics[width=0.95\textwidth]{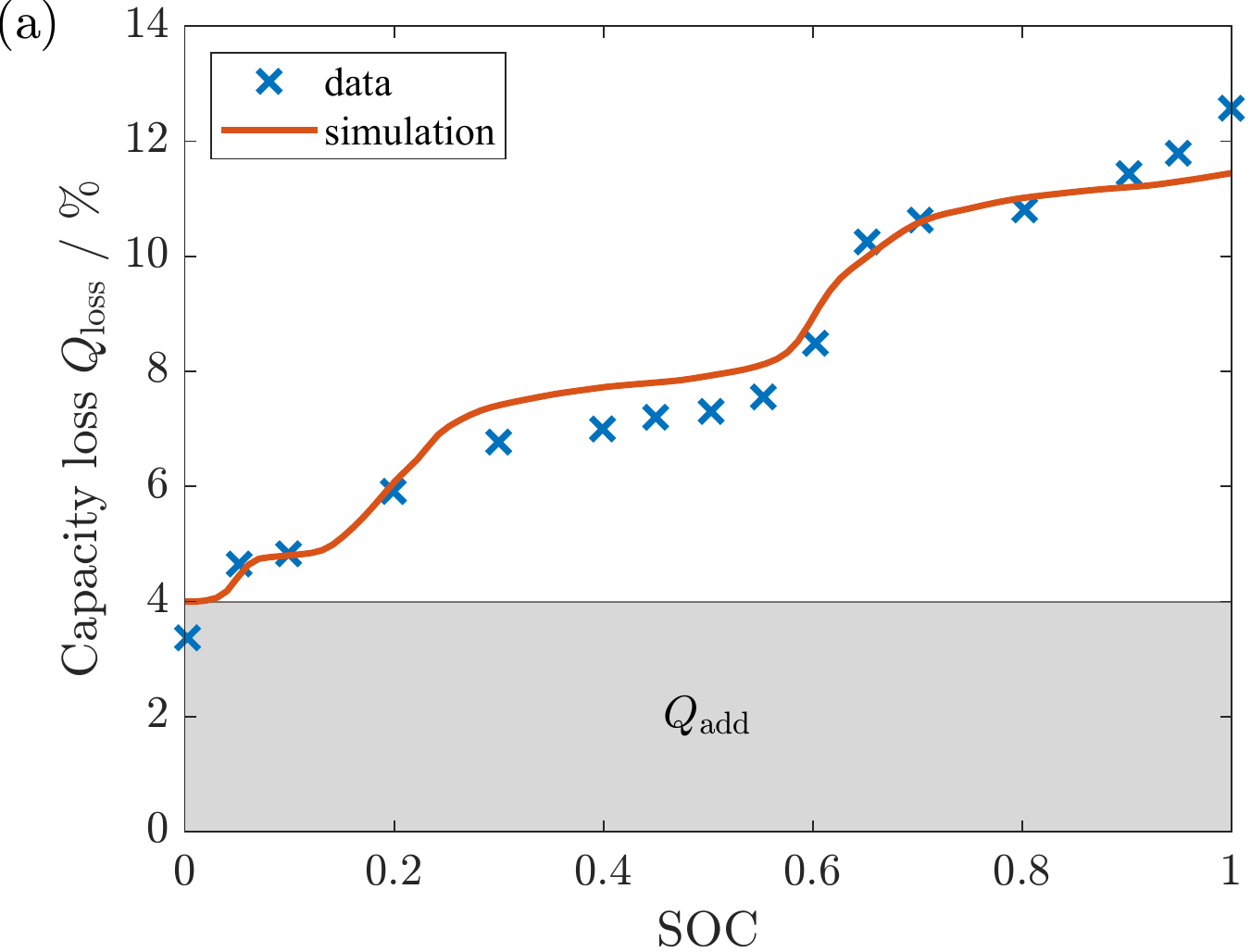}
	\end{minipage}
	\begin{minipage}{0.5\textwidth}
		\includegraphics[width=0.95\textwidth]{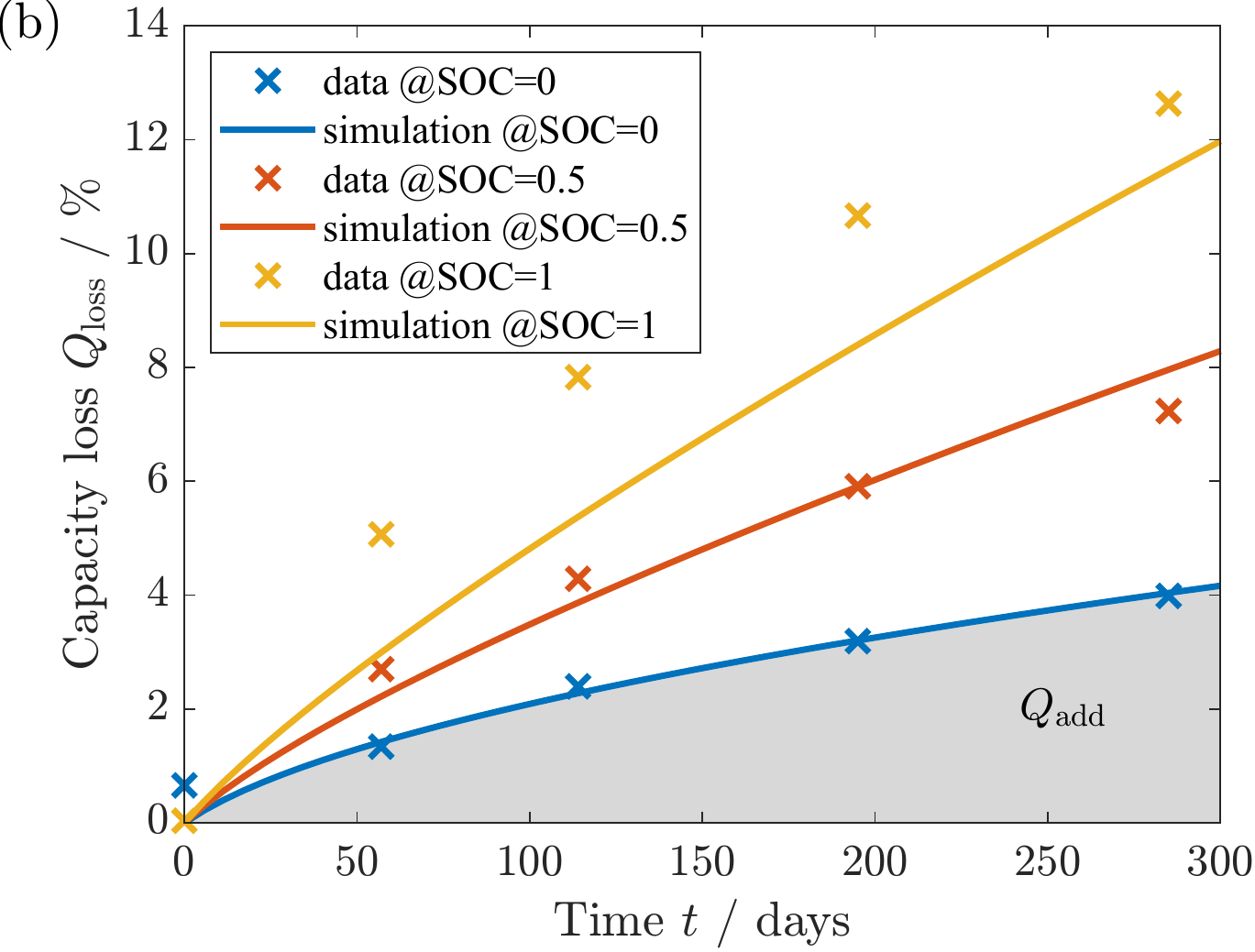}
	\end{minipage}
	\\
	\begin{minipage}{0.5\textwidth}
		\includegraphics[width=0.95\textwidth]{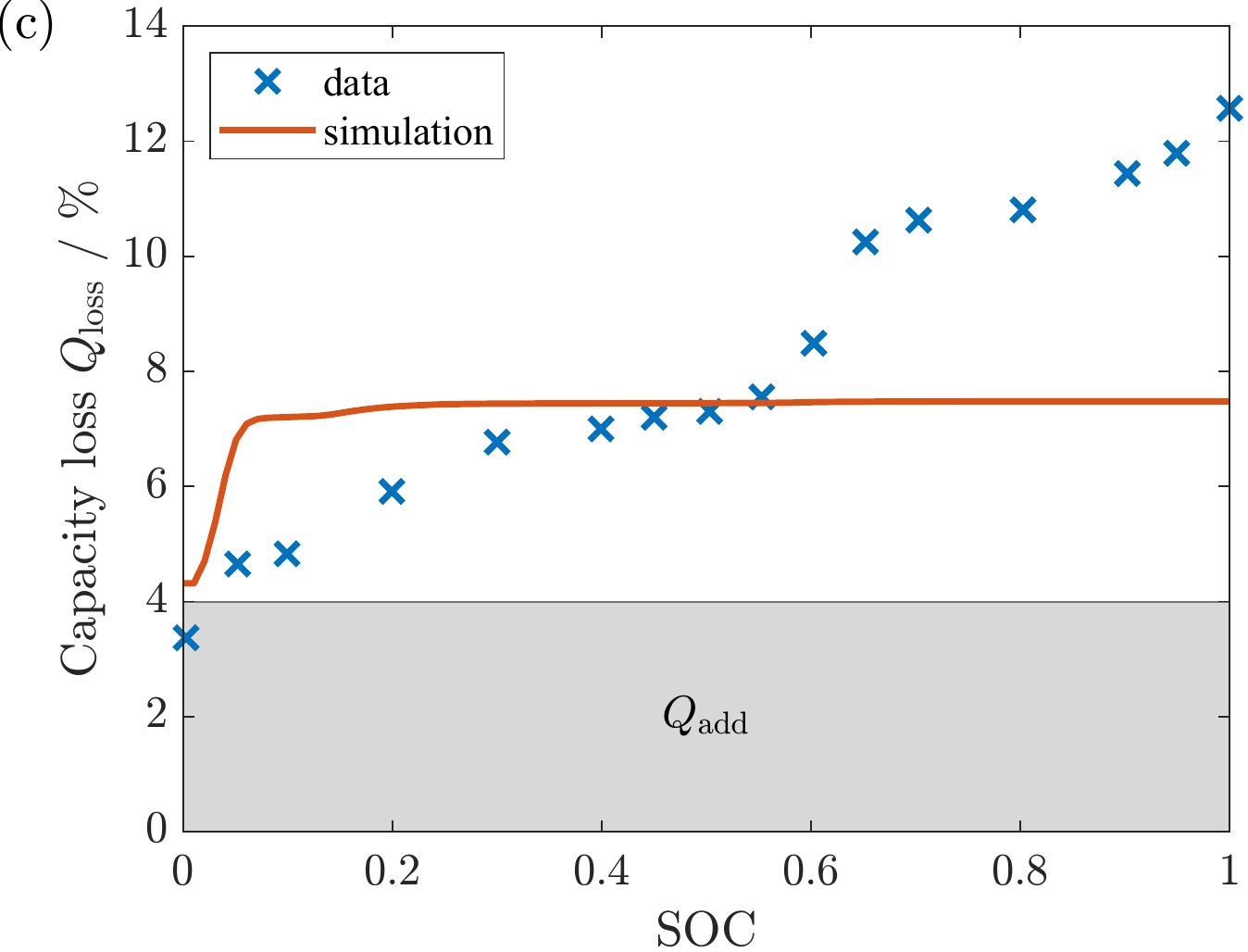}
	\end{minipage}
	\begin{minipage}{0.5\textwidth}		
		\includegraphics[width=0.95\textwidth]{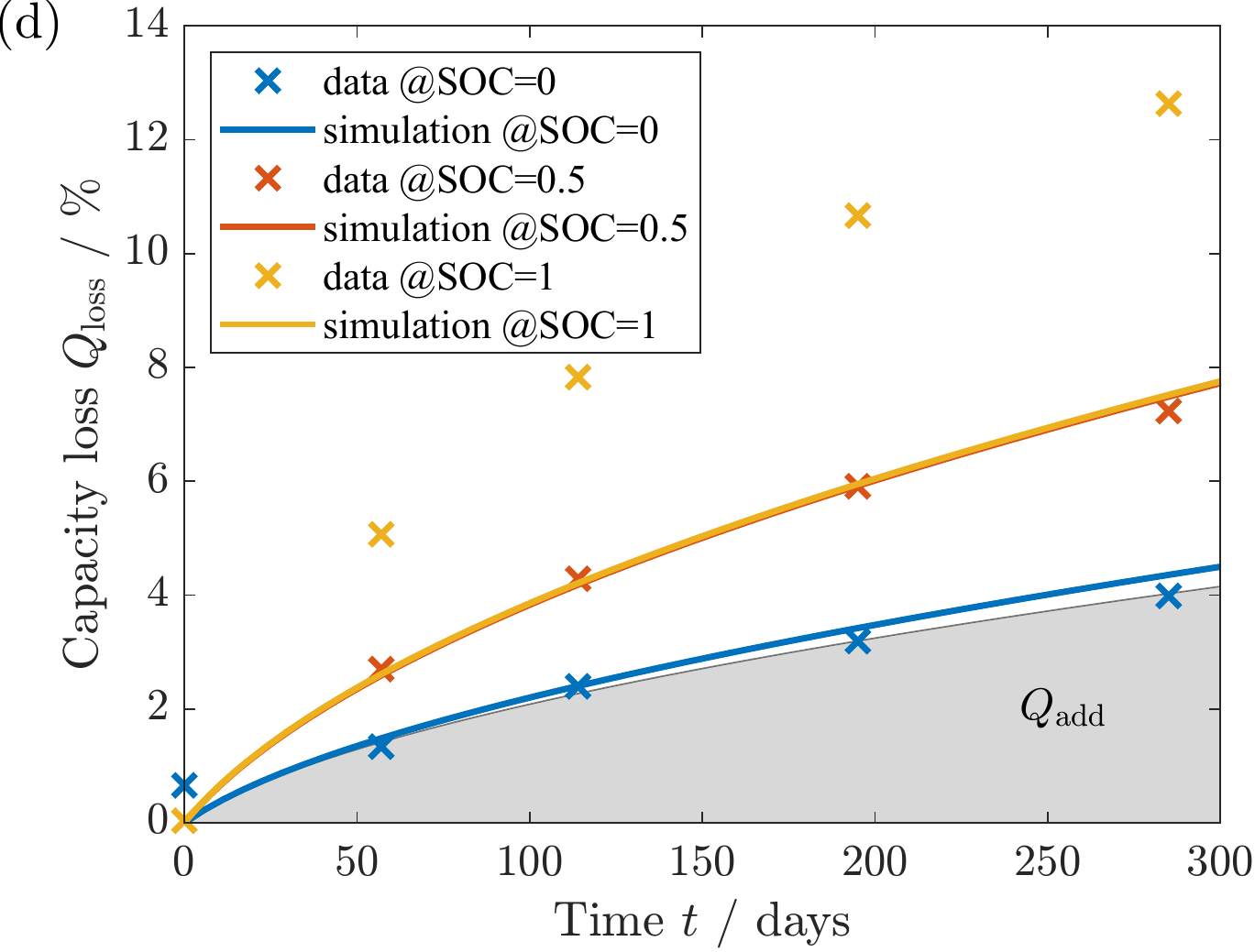}
	\end{minipage}

	\caption{Capacity loss simulated with solvent diffusion in comparison to the data measured in Ref. \cite{Keil2016}. (a) and (b) Fit of the solvent diffusion model to the SOC dependence of capacity loss (a) Capacity loss depending on the SOC. The plot shows a good agreement between simulation and data points. (b) Capacity loss depending on time. Simulation and experiment do not match at medium and high SOC. (c) and (d) Fit of the solvent diffusion model to the time dependence of capacity loss at $\mathrm{SOC}=50\%$ (c) Capacity loss depending on the SOC. The simulation does not show the experimentally observed SOC dependence. (d) Capacity loss depending on time. Excellent agreement is achieved at $\mathrm{SOC}=50\%$ but the simulation does not reproduce the behavior at $\mathrm{SOC}=100\%$.}
	\label{fig:SD-SOC-time}
\end{figure*}

\begin{figure*}[htp]
    % Figure 4
    % width = 2 columns
	\begin{minipage}{0.5\textwidth}
		\includegraphics[width=0.95\textwidth]{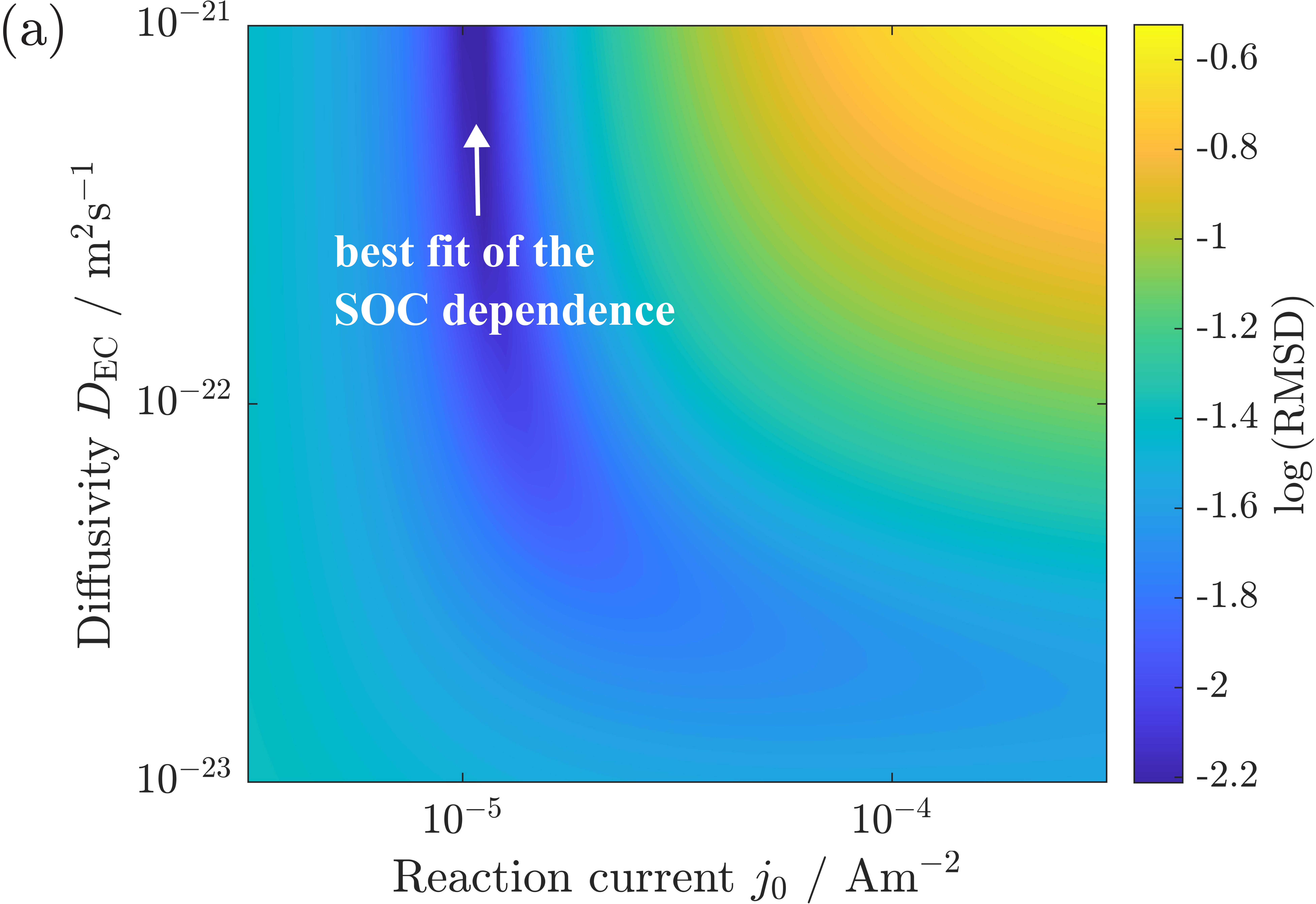}
	\end{minipage}
	\begin{minipage}{0.5\textwidth}
		\includegraphics[width=0.95\textwidth]{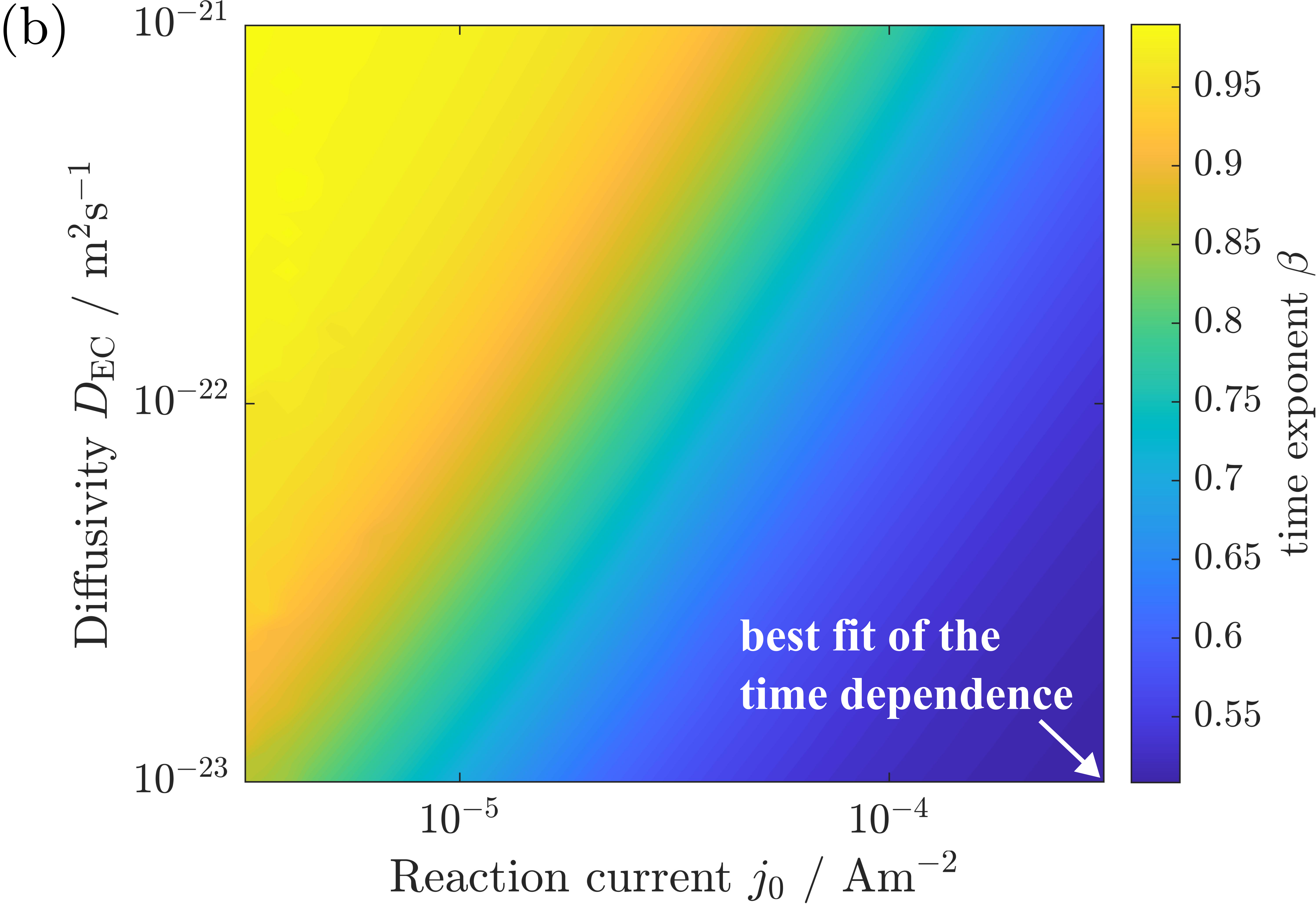}
	\end{minipage}
	\caption{Parameter study for the capacity loss simulated with solvent diffusion with a variation of the reaction current $j_0$ and the diffusion coefficient $D_\mathrm{EC}$. (a) Logarithm of the root mean square deviation (RMSD) of the simulated capacity loss depending on the SOC from the experiment \cite{Keil2016}. The smallest deviation is depicted in dark blue and reached for a small reaction current $j_0$ and large diffusivity $D_\mathrm{EC}$. The best fit of the SOC dependence with the parameters used in \cref{fig:SD-SOC-time}(a) is highlighted. (b) Fitted time exponent $\beta$ to the simulation. The experiment can be approximately described with $t^{0.5}$ depicted in dark blue. Either large values of $j_0$ or small values of $D_\mathrm{EC}$ are needed to achieve a good agreement with $t^{0.5}$. The best fit of the time dependence with the parameters used in \cref{fig:SD-SOC-time}(d) is highlighted.}
	\label{fig:SD-parameter-study}
\end{figure*}

Next, we investigate the solvent diffusion mechanism, which competes with electron diffusion.  
To explain the various experimental observations, our model for solvent diffusion considers the interplay between reaction and transport limitation.
In the following, we set the parameter $\alpha=0.5$, as we observe no qualitative influence of this parameter on our results.
With the same procedure as before, we subtract the data at $\mathrm{SOC}=0$ from the experimental data to examine the SOC dependent capacity fade. 
As above, we analyze the data by Keil et al. \cite{Keil2016} here and provide the analysis for the data by Naumann et al. \cite{Naumann2018} in the supplementary information (see \ref{fig:SD-SOC-time-naumann} and \ref{fig:SD-parameter-study-naumann}).

First, we fit the solvent diffusion model equation (\ref{eq:CL-solvent-diffusion}) to the SOC depending capacity fade data. The corresponding result is shown in \cref{fig:SD-SOC-time}(a). The simulation shows a good agreement with the SOC dependence of the observed capacity loss. The quality of the fit is similar to the one calculated for electron diffusion. This result is contrary to Ref. \cite{Single2018}, where solvent diffusion is only described in the transport limited regime. However, \cref{fig:SD-SOC-time}(b) shows that with the same parameters, the simulation reproduces the time dependence only at small SOC, where the offset from $\mathrm{SOC}=0$ contributes primarily to the capacity loss. At higher SOC, the time dependence does not fit the data points. In this case, the SOC depending capacity fade, which is approximately linear in time, contributes significantly. The linearity of the capacity fade reveals that the simulation is in the reaction limited regime to capture the SOC dependence. In conclusion, matching the SOC dependence to the experiment does not reproduce the time dependence.

In comparison to the fit to the SOC dependence, which revealed the reaction limited regime, we fit the time dependence of the capacity fade next. As we assume to achieve the best accordance at medium SOC, we fit the solvent diffusion model to the capacity loss over time at $\mathrm{SOC}=50\%$. As shown in \cref{fig:SD-SOC-time}(d), we find a perfect agreement of simulation and experiment at the fitted $\mathrm{SOC}=50\%$. In contrast, at high SOC, the simulation gives the same result as at $\mathrm{SOC}=50\%$ and does not capture the experimental data. The deviation at high SOC is explained when regarding \cref{fig:SD-SOC-time}(c). With the parameters obtained from the fit to the time dependence, the simulation shows a constant contribution to capacity loss over a wide range of SOC values. The simulation does not capture the dependence of the capacity loss on the SOC at all. As we observe perfect agreement of the time dependence at $\mathrm{SOC}=50\%$ but no agreement of the SOC dependence, the simulation is in the transport limited regime. Thus, matching the time dependence to the experiment, the simulation does not reproduce the SOC dependence.

%%%%%%%%%%%%%%%%%%%%%%%%%%%%%%%%%%%%%%%%%%%%%%%%%%%%%%%%%%%
% Figure with label fig:SD-SOC-time here !!!
%%%%%%%%%%%%%%%%%%%%%%%%%%%%%%%%%%%%%%%%%%%%%%%%%%%%%%%%%%%

As both fits reveal only one observed behavior, we try to find a possible intermediate regime where the solvent diffusion mechanism captures the SOC dependence and the time dependence with the same parameters. To find this regime, we vary the reaction current and the diffusion coefficient in \cref{eq:CL-solvent-diffusion}. In this parameter study, we compare the root mean square deviation between the simulation and the experimental data points describing the SOC dependence. Additionally, we fit a simple power law $t^\beta$ to the simulated time dependence and evaluate the agreement of the time exponent $\beta$ with the square-root profile, i.e. $\beta=0.5$.

The outcome of the parameter study is depicted in \cref{fig:SD-parameter-study}. On the one hand, \cref{fig:SD-parameter-study}(a) shows the logarithm of the root mean square deviation of the simulation to the experimental data describing the SOC dependence. The smallest values depicted in dark blue refer to the best accordance of simulation and experiment. Therefore, in the given range of parameters, small values of the reaction current $j_0$ and large values of the diffusion coefficient $D_\mathrm{EC}$ are necessary to describe the experimentally observed SOC dependence.

%%%%%%%%%%%%%%%%%%%%%%%%%%%%%%%%%%%%%%%%%%%%%%%%%%%%%%%%%%%
% Figure with label fig:SD-parameter-study here !!!
%%%%%%%%%%%%%%%%%%%%%%%%%%%%%%%%%%%%%%%%%%%%%%%%%%%%%%%%%%%

On the other hand, \cref{fig:SD-parameter-study}(b) depicts the fitted time exponent $\beta$. Here, values around $\beta=0.5$ are required to describe the experimentally observed time dependence, which shows approximately a square-root behavior. The corresponding values are illustrated again in dark blue. From the same range of parameters as before, now large values of the reaction current $j_0$ and small values of the diffusion coefficient $D_\mathrm{EC}$ are needed.

To combine both results, all sets of parameters that describe the SOC dependence do not reproduce the time dependence, and all sets of parameters that describe the time dependence do not reproduce the SOC dependence. The parameter study shows either good accordance with the SOC dependence or good accordance with the $t^{0.5}$ dependence. There exists no intermediate regime, where solvent diffusion reproduces the SOC dependence and the time dependence. As the same experiment exhibits both dependencies, the solvent diffusion mechanism fails to describe the observed SEI growth characteristics. However, solvent diffusion could be one of the reasons for the contribution to the capacity fade that is independent of the SOC.

In this section, we have shown that the diffusion of solvent molecules can not capture the experimentally observed dependencies of capacity fade on SOC and time. An alternative approach is considering additional species like the SEI product or lithium ions inside the SEI, which diffuse through the SEI from the electrode towards the electrolyte \cite{Jin2017,Jana2022}. However, this model requires an explanation of why the SEI product does not precipitate directly at the electrode but diffuses towards the SEI-electrolyte interface instead. Alternatively, it has to be clarified why lithium ions should limit the transport process as they can move fast through the SEI to enable battery operation. The mathematical structure of this particular description would be similar to the model of electron diffusion, as the diffusion of a species from the electrode towards the electrolyte determines the SEI growth rate.

\subsection{Self-discharge: Deviation from square-root behavior}
\begin{figure}[htb]
    % Figure 5
    % width = 1 column
	\centering
	\includegraphics[width=0.45\textwidth]{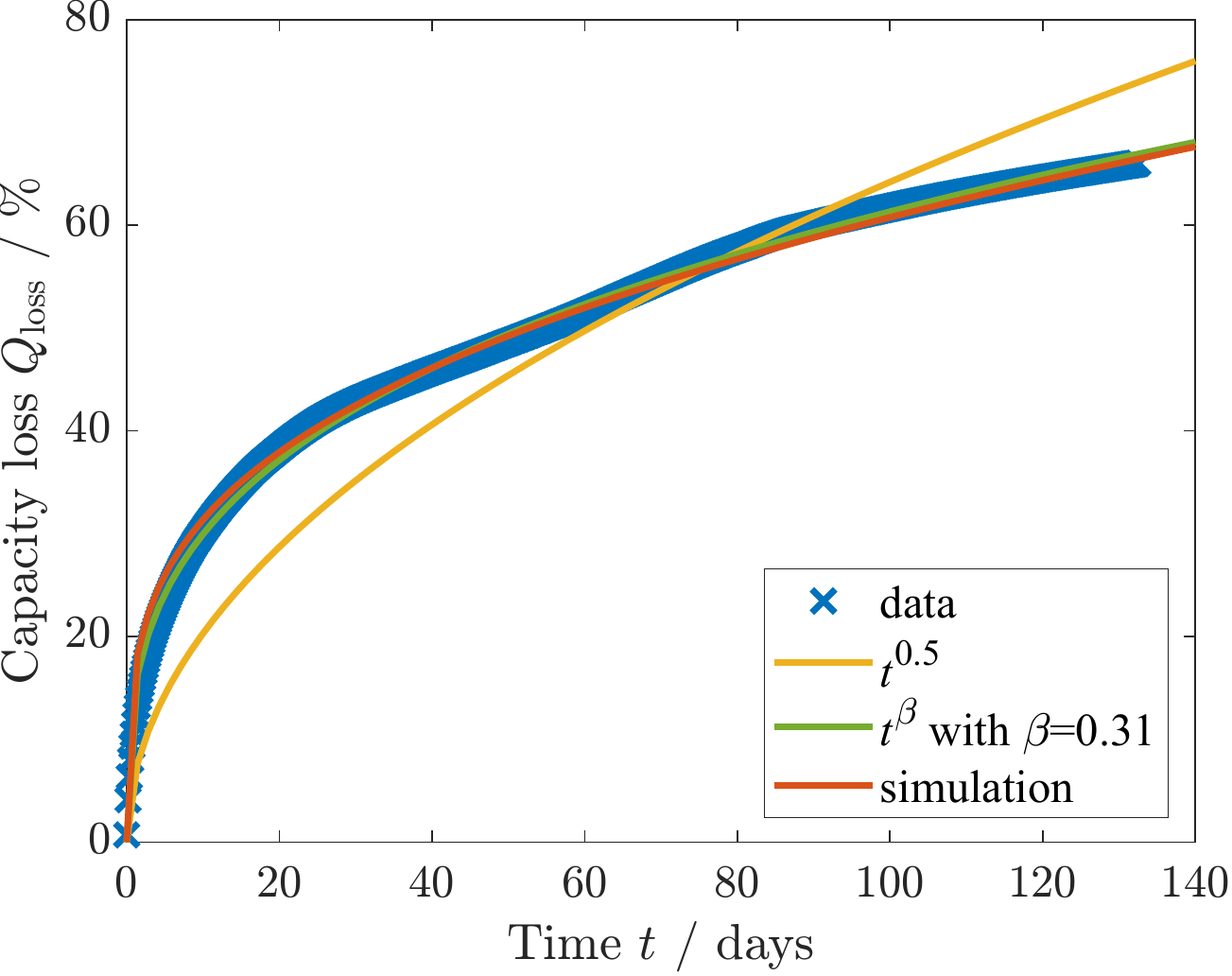}
	\caption{Capacity loss on carbon black during storage. The experimental data from Ref. \cite{Attia2020} show a clear deviation from $t^{0.5}$ behavior. The simulation for electron diffusion can describe the observed time behavior when self-discharge is taken into account.}
	\label{fig:CL_CB}
\end{figure}

So far, we depict the SOC dependence of the capacity loss as a separate feature compared to the time dependence. Now, we want to highlight the interaction between the SOC dependence and the $t^{0.5}$ time dependence due to self-discharge. For a good demonstration of this effect, we refer to an experiment by Attia et al. measuring a particularly large capacity loss on carbon black over time \cite{Attia2020}. The experimental data are shown in \cref{fig:CL_CB}.

For this experiment, Attia et al. reviewed the square-root behavior in time and fitted power laws $t^\beta$ to the experimentally observed capacity fade \cite{Attia2020}. For storage, they observed apparent time exponents lower than $0.5$, e.g. $\beta \sim 0.3$ as depicted in \cref{fig:CL_CB}, in contrast to the usual $t^{0.5}$ behavior expected for transport limitation. However, this can not be explained by leaving the transport limited regime, as values larger than $0.5$ indicate reaction limitations \cite{Kolzenberg2020}.

%%%%%%%%%%%%%%%%%%%%%%%%%%%%%%%%%%%%%%%%%%%%%%%%%%%%%%%%%%%
% Figure with label fig:CL_CB here !!!
%%%%%%%%%%%%%%%%%%%%%%%%%%%%%%%%%%%%%%%%%%%%%%%%%%%%%%%%%%%

As discussed before, the growth rate of the SEI depends on the anode voltage via the SOC calculated with respect to the initial capacity. However, the SOC can not be assumed constant in time but gets effectively reduced by the amount of charge consumed due to SEI growth. This effect of self-discharge leads to a deviation of the simulated SEI growth from a simple square-root behavior in time, particularly for large capacity losses.

In contrast to the former sections, we can not determine the contribution to the capacity loss independent of the SOC separately, as Ref. \cite{Attia2020} provides no data at $\mathrm{SOC}=0$. However, we assume that both SOC dependent and independent parts can contribute significantly to the total capacity loss. Our assumption is supported by results from cryogenic electron microscopy of the SEI on carbon black \cite{Huang2019}. This study reports a thin and a thick SEI layer on different particles. In addition, Ref. \cite{Jana2022} describes two SEI growth modes, namely homogeneous and heterogeneous SEI growth. Therefore, it is reasonable to assume that two different mechanisms may lead to the observed SEI thickness. Due to our former results, we attribute the SOC dependent term to our electron diffusion model (\ref{eq:CL-electron-diffusion}) and the SOC independent term to an additional mechanism. Solvent diffusion would be a candidate for this mechanism.

As motivated, we divide the total capacity loss into a contribution with nontrivial SOC dependence and a contribution with trivial SOC dependence. Here, we set $Q_\mathrm{SEI,0} = 0$ to reduce the number of parameters. \cref{fig:CL_CB} illustrates the simulation of the capacity loss. It matches the experiment reasonably with similar quality as the simple power law fit. Therefore, we highlight that electron diffusion can well describe SEI growth as a transport limited process, i.e. $t^{0.5}$ behavior for constant voltage, when carefully considering any change in the SOC and the voltage $U(\mathrm{SOC})$, respectively. Considering this self-discharge, our model for electron diffusion is able to describe deviations from the $t^{0.5}$ dependence. Our results reveal that smaller apparent exponents of the power law $t^\beta$, as observed in Ref. \cite{Attia2020}, are still consistent with transport limited SEI growth models. Instead, exponents larger than $0.5$ imply the transition to the reaction limited regime \cite{Kolzenberg2020}.

\section{Conclusions}
We have investigated the SOC and time dependence of capacity fade during calendar aging and compared the electron diffusion with the solvent diffusion mechanism. The relevant transport mechanism leading to SEI growth has to capture the SOC and the time dependence with the same parameters.

We have found that the electron diffusion mechanism can reproduce the SOC dependence and the time dependence of the capacity loss simultaneously. We conclude that electron diffusion is able to describe the characteristics of the SOC dependent capacity loss. Only at very low and high SOC do additional degradation effects not captured in our model lead to deviations. Their interplay deserves further investigation.

In contrast, we have shown that solvent diffusion can reproduce either the SOC dependence or the time dependence of capacity fade. Our parameter study approves that there is no intermediate regime, which captures both dependencies with the same parameters. Thus, solvent diffusion fails to describe the observed features of the SOC dependent capacity loss.

Finally, we have emphasized the importance of the interaction between SOC dependence and time dependence of capacity fade. We have demonstrated that our model for electron diffusion captures experimentally observed deviations from the usual $t^{0.5}$ law by considering the effect of self-discharge. Consequently, smaller apparent time exponents, as reported in Ref. \cite{Attia2020}, do not contradict transport limited SEI growth. In contrast, larger exponents may indicate reaction limitations \cite{Kolzenberg2020}.

To conclude our study, we have shown that only the electron diffusion mechanism can explain the SOC and time dependence simultaneously. This mechanism can even reproduce capacity fade experiments that deviate from the square-root behavior in time. In agreement with Refs. \cite{Jana2022,Huang2019}, we have divided the total capacity loss into two contributions. However, the reason for the SOC independent capacity loss needs further investigation.

To take future calendar aging studies into account, we propose to appropriately state experimental data of capacity loss as a function of SOC and time. The check-up cycles have to be done carefully to reduce their influence on the self-discharge effect, i.e. it is important to state to which SOC the cells are charged after the check-ups. Additionally, it is crucial to provide the anode open circuit voltage curve for the investigated cells to analyze the SOC dependence of SEI growth.

\section*{Acknowledgments}
This work was funded by the German Research Foundation (DFG) in the framework of the research training group SiMET – Simulation of Mechano-Electro-Thermal Processes in Lithium-Ion Batteries (281041241/GRK 2218).

\bibliography{refs}
\bibliographystyle{elsarticle-num}

\end{document}